\documentclass[a4paper]{mem}
\usepackage{natbib}
\usepackage{graphicx}
\def\ML{\mbox{$M/L$}}
\usepackage[a4paper]{hyperref}
\idline{73}{23}
\begin{document}
   \title{Galaxy dynamics with the Planetary Nebula Spectrograph
}

   \author{ N.R. Napolitano\inst{1}, A.J. Romanowsky\inst{2}, N.G. Douglas\inst{1}, M. Capaccioli\inst{3}, M. Arnaboldi\inst{4} K. Kuijken\inst{1,5}, M.R. Merrifield \inst{2}, K.C.  Freeman\inst{6}, O. Gerhard\inst{7}}

   \offprints{N.R. Napolitano}
\mail{nicola@astro.rug.nl}

   \institute{Kapteyn Astronomical Institute, Groningen\\ 
              \and School of Physics \& Astronomy, University of Nottingham \\
              \and INAF -- Observatory of Capodimonte, Naples\\
              \and INAF -- Observatory of Pino Torinese, Turin\\
              \and University of Leiden\\
	      \and RSAA, Mt. Stromlo Observatory \\
         \and Astronomical Department, University of Basel
             }

   \abstract{ The Planetary Nebula Spectrograph is a dedicated instrument for measuring radial velocity of individual Planetary Nebulae (PNe) in galaxies. This new instrument is providing crucial data with which to probe
the structure of dark halos in the outskirts of elliptical galaxies in particular, which are traditionally lacking of easy interpretable kinematical tracers at large distance from the center. Preliminary results on a sample of intermediate luminosity galaxies have shown little dark matter within 5~$R_{\rm eff}$ implying halos either not as massive or not as centrally concentrated as CDM predicts \citep{ral03}. We briefly discuss whether this is consistent with a systematic trend of the dark matter content with the luminosity as observed in an extended sample of early-type galaxies.

   }
   \authorrunning{N.R Napolitano et al.}
   \titlerunning{Galaxy dynamics with the PN.S}
   \maketitle
%

\section{Introduction}
The kinematical information of early-type systems has been historically made difficult by the intrinsic lack of easy interpretable kinematical tracers like cold gas in spirals. 
So far, the main knowledge of their kinematics was constrained in the inner regions (R$\le$ 2 R$_e$, where R$_e$ is the effective radius which contains only half of the total galaxy light) through the standard technique based on the spectra of the integrated light. For this reason, much of the interesting information concerning the outer radii was inaccessible.
Planetary Nebulae (PNe) have been proven to be excellent tracers for the dynamics of outer regions of early-type galaxies. Through their powerful [OIII] emission at 5007 \AA, and 4959 \AA, they are easily detectable and their radial velocity measurable also in halo regions of distant galaxies. PNe discrete radial velocity fields have been recognised as an efficient tool to study the kinematics of the halo regions of ellipticals in a large sample of observational programs \citep{1,2,22,14,32,ral03,peng04}.\\ 
PNe are the natural candidates to gauge the stellar kinematics in the outskirts of ellipticals. Their kinematics can be combined together with the absorption-line spectra of the inner regions to obtain a picture of a galaxy's stellar dynamics from its center to several R$_e$, derive mass and angular momentum estimates and search for kinematical signatures of interactions.\\ 
So far, the small PNe number statistics did not allow to draw definitive results about galaxy formation theories and dark-matter content in early-type galaxies, but the perspectives of the PNe science is triggering a strong effort to develop new instruments and observing techniques. \\
In July 2001 a dedicated instrument to the PN kinematics in galaxy systems, the {\bf Planetary Nebula Spectrograph} (PN.S), was
commissioned at the 4m Herschel Telescope on La Palma. Here we discuss preliminary results on a sample of intermediate luminosity galaxies. These galaxies shown little dark matter within 5~$R_{\rm eff}$ as well as a very low angular momentum at the same distances: these results seem to conflict with expectation from the $\Lambda$CDM cosmology. 

   \begin{figure}
   \centering
   \resizebox{\hsize}{!}{\rotatebox[]{0}{\includegraphics{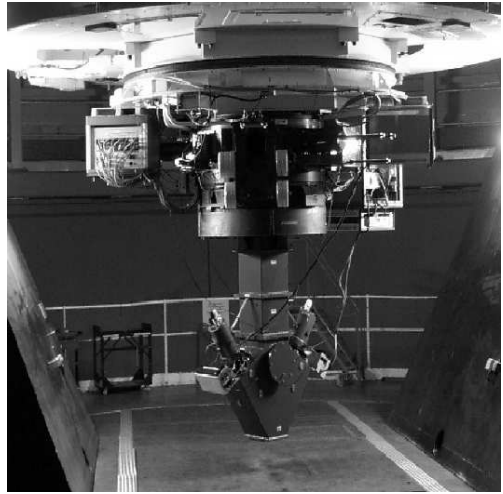}}}
   \caption{\small The PN.S at the Cassegrain focal station of the
William Herschel Telescope, which has built-in spectral
lamps for calibrating the measured velocities. The PNS measures
CDI images simultaneously by means of the two detectors 
shown here (photo courtesy of R.A. Hijmering).}
              \label{Fig1}%
    \end{figure}

   \begin{figure*}
   \centering
   \resizebox{\hsize}{!}{\rotatebox[]{0}{\includegraphics{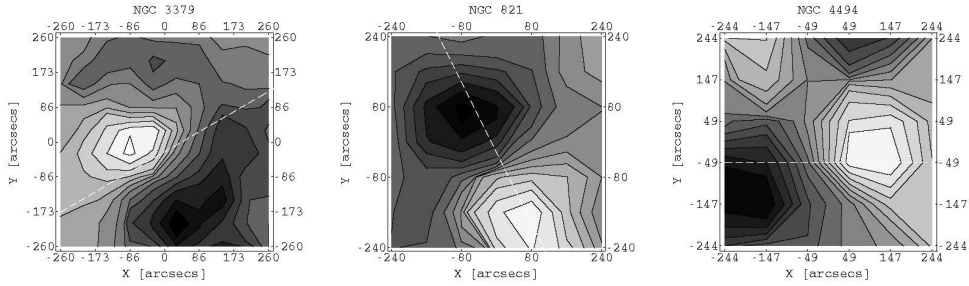}}}
   \caption{\small  Smoothed PNe velocity field from PN.S data, 
darker to brighter regions are negative to positive radial velocities. Maximum velocity is $V_{\rm max}=65, 80, 55 km s^{-1}$ for NGC~3379, NGC~821 and NGC~4494 respectively (see text). 
Dashed line is the P.A. of the inner stellar major kinematical axis. 
Misalignment of the PNe axis with respect the inner stellar kinematics, 
kinematical substructures and twist of the isovelocity contours are evident.
}
              \label{Fig2}%
    \end{figure*}

\section{The PN.Spectrograph}
The PN.S was specifically designed for the measurement of kinematics of
extragalactic planetary nebulae. It allows this type of work to be
carried out a factor of ten more efficiently than possible so far, due
to a combination of very efficient optics optimized for the dominant
emission line of PNe spectra, a wide field, and most importantly a
design which allows the PNe to be discovered and their spectra
measured in a single observation (which means it is no longer
necessary to make two separate observing campaigns, usually in
different seasons, before kinematics can be obtained.)
The PN.S incorporates two highly efficient slitless spectrographs
back-to-back (Fig. 1).  Via a common [OIII] filter this results in simultaneous
images of the field in which the PNe are easily distinguished from
stars.  Their velocities can be derived from the relative shifts in
their positions in the two images.  This ``Counter-Dispersed Imaging"
technique is described in \citet{dou02}, along with other details
of the project. Its value lies in the high optical efficiencies attained, in
the absence of any astrometric requirements, and in the fact that it 
is a single-step procedure. 
Our first project, currently in progress, 
is to examine the dark matter halos of
apparently round, moderate-luminosity elliptical galaxies, too
small to have been easily addressed by X-ray or gravitational-lensing
analysis.

\section{PN Spectrograph survey: first results}
The primary program of the PN.S is to survey a dozen bright ($m_B\leq$~12), round (E0--E2), nearby ($D\la20$~Mpc) ellipticals,
obtaining 100--400 PN velocities in each.
Based on the statistical distribution of galaxy shapes
\citep{lam92},
we estimate that by selecting for projected axis ratios $q\ga0.8$, 75\% of our sample 
will have intrinsic gravitational potential ellipticities $\epsilon_\Phi \la 0.1$
and thus can be well characterized by spherical dynamical models.
Other than the constraints above, our sample is designed to encompass a broad spectrum
of elliptical galaxies, 
with a large range of stellar light parameters (luminosity, concentration, shape),
rotational importance, and environment.

Our program has so far been beset by bad weather,
but we have obtained extensive data on the galaxies
NGC~821, NGC~3379, and NGC~4494, three ordinary intermediate luminosity galaxies ($\sim$$L^*$).
For each we have
an initial data set of $\sim$~100 PN velocities out to $\sim$~5~$R_{\rm eff}$,
but with additional data reduction we expect these to increase to $\sim$~200 each.

{\em Velocity structure} -- In Fig.1 we show a gaussian smoothed version of the PN radial velocity fields of the galaxies \citep{ral03}. In case of NGC~3379 and NGC~4494, there is a tilt of the major kinematical axis of the PNe sample with respect to the stellar long-slit data. 
On the contrary, a quite good alignment is found for NGC~821 as well as it has been found also in earlier studies \citep[NGC~4697:][]{29}. 
Incidentally, the latter are viewed quite edge-on, while NGC~3379 and NGC~4494 are possibly nearly face-on. 
In all these systems, the rotation velocity curves \citep[obtained following][]{30}  increase in the inner parts (up to $\sim 2~R_{\rm e}$) and rapidly drop in the outer parts 
down to zero at the outermost data points. Namely, for NGC~3379 PNe have maximum rotation velocity, $V_{\rm max}=65\pm30$ km s$^{-1}$, at R=70$''$ ($\sim 2 R_{\rm e}$) and then they drop in velocity to $V_{\rm min}=10\pm30$ km s$^{-1}$ at R=165$''$ ($\sim 4.7 R_{\rm e}$); for NGC~821 we measured $V_{\rm max}=80\pm30$ km s$^{-1}$, at R=75$''$ ($\sim 1.5 R_{\rm e}$) and $V_{\rm min}=20\pm30$ km s$^{-1}$ at R=150$''$ ($\sim 3 R_{\rm e}$); for NGC~4494 we found $V_{\rm max}=55\pm20$ km s$^{-1}$, at R=70$''$ ($\sim 1.4 R_{\rm e}$) and $V_{\rm min}=20\pm30$ km s$^{-1}$ at R=100$''$ ($\sim 2 R_{\rm e}$).
This conflicts with major merger simulations that predict rapid outer rotation: $v/\sigma \ga$~1 outside 2~$R_{\rm eff}$ \citep{wehe96,benba00}.\\
Is this an indication of some secondary evolution processes in some of 
these galaxies (i.e., tidal interactions or merging), 
or is this the result of different stellar population inhabiting inner (disk population) and outer (bulge/halo population) 
regions, which appear to dominate depending on the viewing angle?

{\em Mass distribution} --  The three galaxies considered here, as well as NGC~4697, velocity dispersions decline rapidly with radius (see Fig. 2).
Simple Jeans models with a moderate degree of anisotropy indicate 
total masses consistent with the visible stars only:
a benchmark parameter $\Upsilon_{B\rm 5}$ ($M/L$ inside 5~$R_{\rm eff}$) is
$\sim$~6--15 \citep{ral03},
while stellar populations should have $\Upsilon_B \sim$~3--12 \citep{18}.
For NGC~3379, we have constructed more versatile orbit models
to allow for the infamous mass-anisotropy degeneracy,
and to extract as much information as possible out of the discrete PN velocity data \citep{rk01}.
These tightly constrain $\Upsilon_{B\rm 5}$ to be $7.1\pm0.6$.
There are still systematic uncertainties in this study,
notably the possibility that the galaxy contains a component that is flattened along the line-of-sight;
however, independent confirmation of a low $M/L$ comes from an HI ring \citep{sch89},
and from a steeply declining dispersion in the GCs \citep{beas04} ---which
are unlikely to reside in a flattened system.

\section{Dark-to-luminous properties of early-type galaxies}
Beside intrinsic indeterminacies in modeling, one can point out that these intermediate luminosity galaxies ($L\sim L_*$) are somehow peculiar systems.
   \begin{figure}
   \centering
   \resizebox{\hsize}{!}{\rotatebox[]{0}{\includegraphics{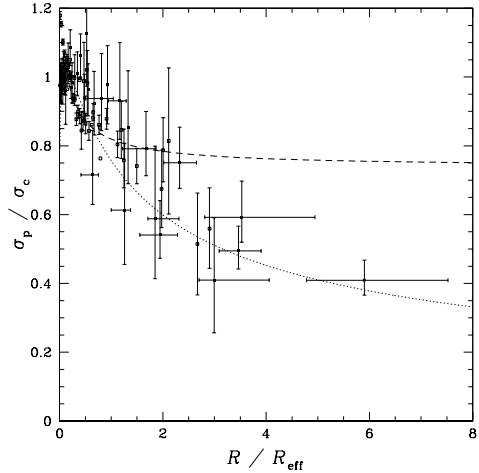}}}
   \caption{\small Projected velocity dispersion profiles for four elliptical galaxies
(see text)
scaled and stacked, as a function of radius, in units of $R_{\rm eff}$.
 Open points show PN data; solid points show long-slit stellar data.
 Simple model predictions are shown for comparison:
 a singular isothermal halo (dashed line) and a constant-$M/L$ galaxy (dotted line).}
              \label{Fig3}%
    \end{figure}
Their structural and kinematical properties make them a different class 
with respect to the brightest systems. Indeed, from the photometric point 
of view, bright galaxies have usually boxy isophotes and flat inner cores 
in the light distribution while fainter galaxies are mostly disky with 
power-law inner light profiles \citep{35,668,fab97}.
Recently, some lines of evidence has been found about the persistence 
of such a dichotomy also in the dark matter properties of these systems.
   \begin{figure*}
   \centering
   \resizebox{\hsize}{!}{\rotatebox[]{0}{\includegraphics{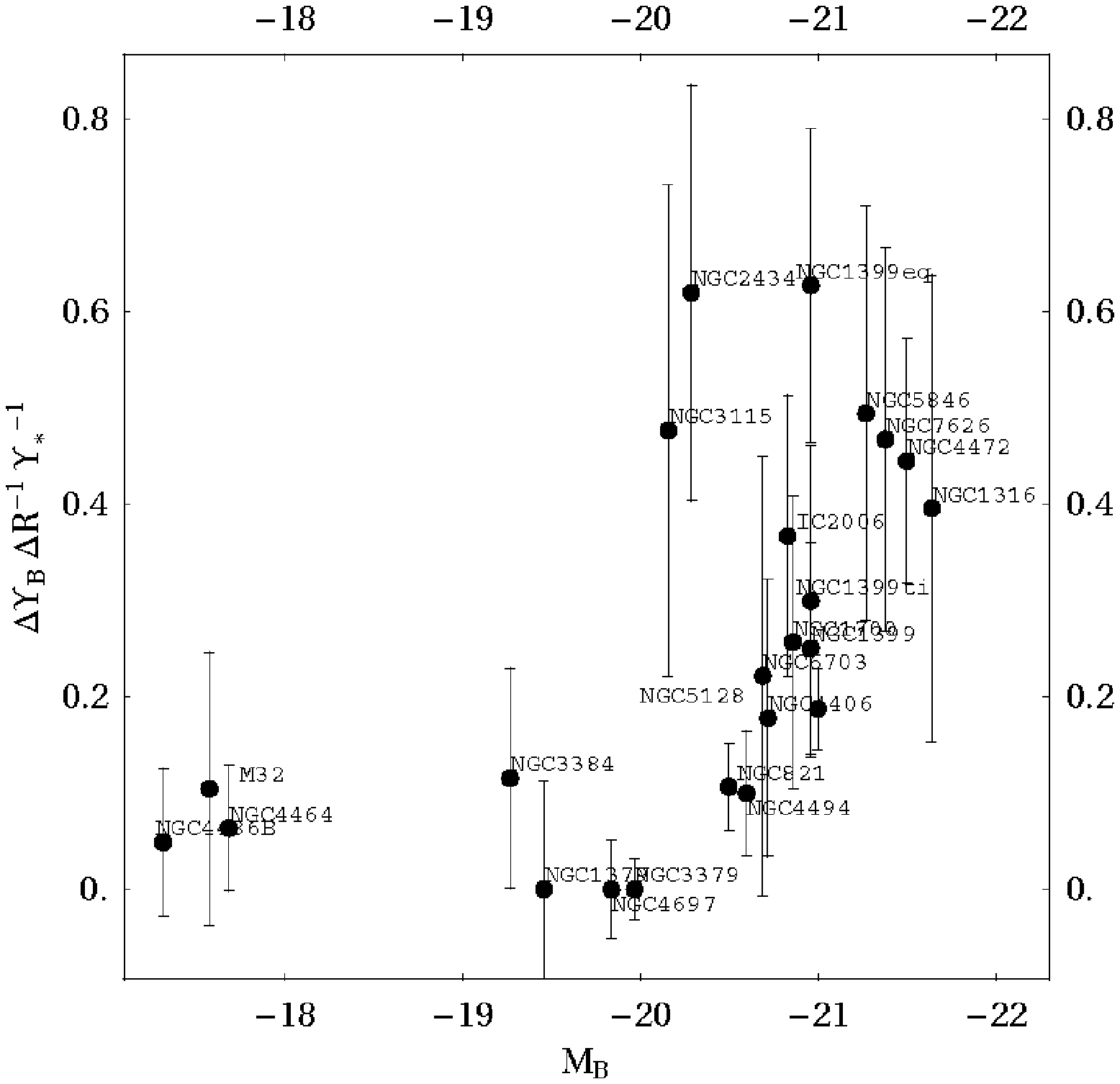}}
   \includegraphics{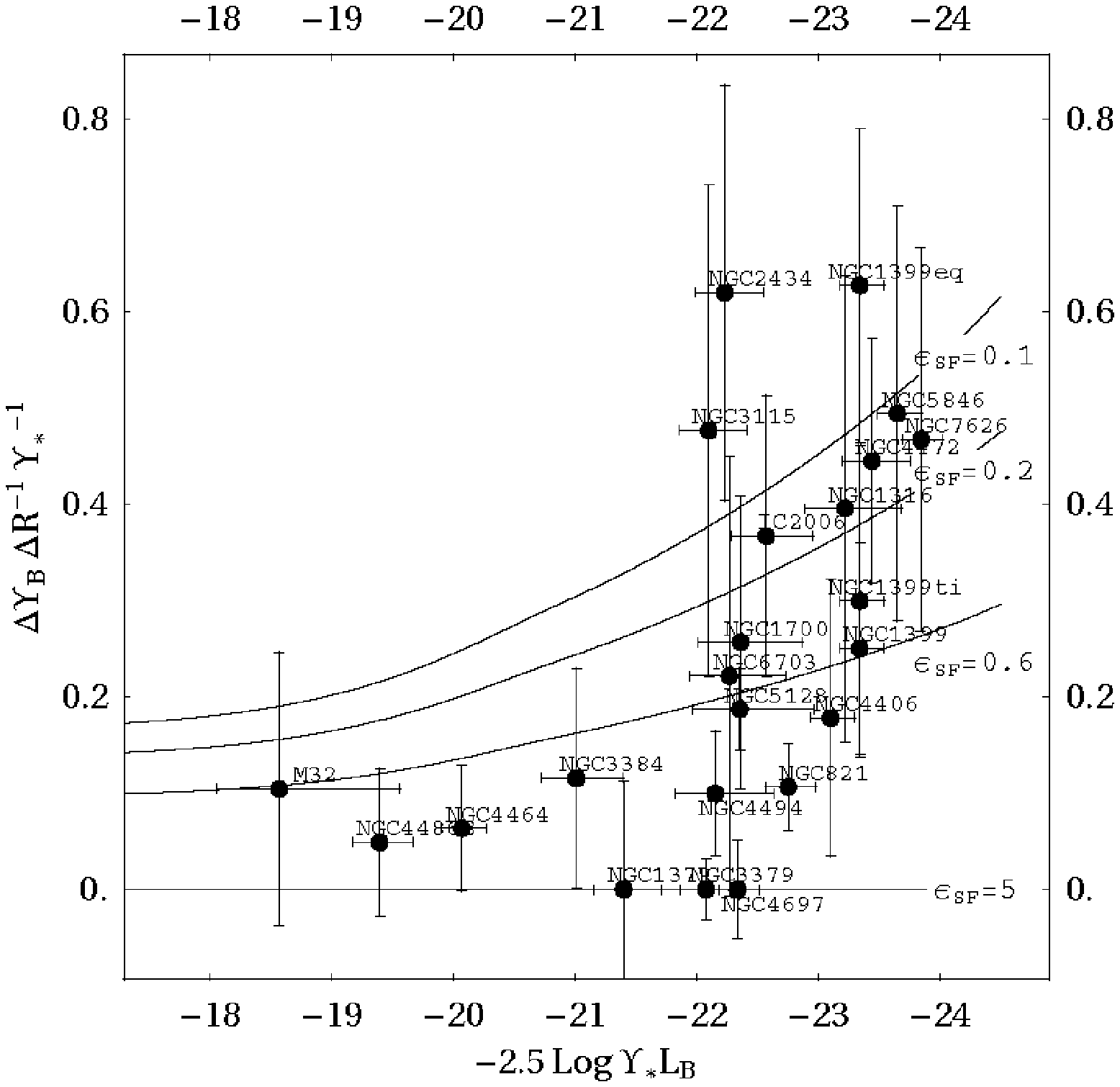}}
   \caption{\small Left: \ML\ gradients versus total luminosity. Right: \ML\ gradients versus total stellar mass, $M_{\rm{l,tot}}=\Upsilon_* \times L_{B,\odot}$. Solid curves are prevision from $\Lambda$CDM parametrised with different stellar formation efficiencies. The observed \ML\ trends with luminosity (and then with $M_{\rm{l,tot}}$) are contemplated in the $\Lambda$CDM framework, but low luminosity systems seem to require an unrealistic star formation efficiency (see text).}
              \label{Fig3}%
    \end{figure*}

\citet{888,nap03} have discussed the correlation between the structural parameters and the radial trend in the
mass-to-light ratios (M/L) using archive kinematic data up to 
$\sim 6 R_{\rm e}$ for an heterogeneous sample of ellipticals.
The radial trend of \ML\ can be quantified by the \ML\ gradients defined by
\begin{equation} 
\scriptsize
\frac{\Delta\Upsilon}{\Delta r}=\frac{\Upsilon_{\mathrm{2}}-\Upsilon_{\mathrm{1}}}{r_{\mathrm{out}}-r_{\mathrm{in}}}=
\frac{\Upsilon_*}{\Delta r}\left[ 
\left(\frac{M_{\rm d}}{M_{l}} \right)_{\mathrm{2}} - \left(\frac{M_{\rm d}}{M_{l}} 
\right)_{\mathrm{1}}\right] ,
\label{grad0} 
\end{equation}
where $M_{l}$ and $M_{\rm d}$ are respectively the luminous (stellar) and the dark masses enclosed within a certain radius (2=outermost, 1=innermost radius), $\Upsilon_*$ is the \ML\  of the galaxy stellar population (it can be computed using spectral information of the galaxy light and stellar systesis models). 
This is the absolute value of the gradient, but to normalise galaxies of different mass and size to
similar scales, we compute
\begin{equation} 
\scriptsize
\frac{R_e {} \Delta\Upsilon}{\Upsilon_*~\Delta r}=\frac{R_e}{\Delta r}\left[ 
\left(\frac{M_{\rm d}}{M_{l}} \right)_{\mathrm{2}} - \left(\frac{M_{\rm d}}{M_{l}} 
\right)_{\mathrm{1}}\right].
\label{grad}
\end{equation}

We will adopt in the following the notation $\delta \Upsilon_{R,\Upsilon_*} =(R_e/ \Delta\Upsilon )/( \Upsilon_*\Delta r)$. As $M_d/M_{l}$ can be reasonably assumed to be a radially linear quantity, i.e. without strong local gradients, $\delta \Upsilon_{R,\Upsilon_*}$ allows us to evaluate the
relative trend in the radial \ML\ distribution among galaxies in our
catalog independently on the radial coverage of the \ML\ estimates, under the assumption of constant $\Upsilon_*$.
Roughly speaking, galaxies with preatty constant \ML\ like the one discussed in the previous paragraph should have \ML\ gradients nearly equal to zero. Galaxies with increasing \ML\ due to extra mass residing in the halo, should show non-zero gradients, as much larger as larger is the amount of dark matter. In Fig. 4 (left), for instance is the \ML\ gradients against the total luminosity for a sample of 21 galaxies \citep[see][for further details]{nap03}.\\  
We have checked if this trend with luminosity is actually consistent with the expectations from the $\Lambda$CDM. In order to do that we build spherical representations of early-type galaxy mass profiles, using a constant-$M/L$
model for the stellar distribution plus a $\Lambda$CDM model of the dark halo \citep[][and reference therein]{bull01}.
We combine these components and derive the \ML\ predictions for various model parameters. From eq. 2, it is easily understood that this quantitity is strongly dependent on the ratio between the total dark mass in the halo, $M_{\rm{d,tot}}$, and the total stellar mass, $M_{\rm{l,tot}}$, which can be related to the net star formation efficiency $\epsilon_{\rm{SF}}$ (including stellar mass loss):
\begin{equation}
f_{\mathrm{b}}^{-1} \approx \frac{M_{\rm vir}}{M_{\mathrm{b}}}= 1+\frac{\epsilon_{\rm{SF}}~M_{\rm d,tot}}{M_{\rm l,tot}} ,
\end{equation} 
where $M_{\mathrm{b}}$ is the baryonic mass, $f_{\mathrm{b}}$ =0.17 (Bennett et al. 2003) is the cosmic baryon fraction, and we used $M_{\rm vir}=M_{\mathrm{b}}+M_{\rm d,tot}$ and $M_{\rm l,tot}=\epsilon_{\rm{SF}}M_{\mathrm{b}}$. In this way we can parametrise the modeled \ML\ with the star formation efficiency. The latter does not vary arbitrarly in galaxies and recent estimates place its limit between 0.1 and 0.6 in early type galaxies. In Fig. 4 (right), the observed and modeled gradients are shown. There seems to be a clear dichotomy:
galaxies with $M_{\rm l,tot} < 1.4\times10^{11} M_{\odot}$ ($M_{\rm B}\sim -20.4$) are generally consistent with
$\epsilon_{\rm SF} > 0.6$ which violates the observed baryon budget, while galaxies with $M_{\rm l,tot} > 1.4\times10^{11} M_{\odot}$  are all consistent with $\epsilon_{\rm SF}$ between 0.1 and 0.6. Such a dichotomy reminds to previous evidences found with respect galaxy structural properties \citep{888,nap03,cap91,cap92a}, clearly addressing a connection with galaxy formation mechanisms. In particular more investigations are needed in order to solve the conflict raised by the intermediate/low luminosity systems, where the critical ingredients could be the traitment of baryon physics and feedbacks processes which could have affected the dark matter distribution and caused the discrepancies with the predicted properties by collisionless N-body simulations.

\begin{acknowledgements}
NRN is receiving grant from the EU within the FP5 program (Marie Curie fellowship).
\end{acknowledgements}

\bibliographystyle{aa}

\end{document}